\documentclass[12pt]{article}

\usepackage{latexsym}                            
\usepackage{amsfonts}                            
\usepackage{amssymb}                             
\usepackage{amsmath}                             
\usepackage{slashed}
\usepackage{cancel}

\usepackage{graphicx}
\usepackage{hyperref}

\newcommand\pubnumber{RBRC-1292}
\newcommand\pubdate{\today}

\newcommand{\Tr}{{\mathrm{Tr}}}

\newcommand{\alfive}{{\alpha_5}}

\newcommand{\alfivehat}{{\hat\alpha_5}}

\newcommand{\CP}{{CP}}
\newcommand{\CPbar}{{\overline{CP}}}
\newcommand{\CPviol}{{\cancel{CP}}}
\newcommand{\MSbar}{{\overline{MS}}}

\newcommand{\mcC}{{\mathcal C}}
\newcommand{\mcL}{{\mathcal L}}
\newcommand{\mcO}{{\mathcal O}}
\newcommand{\mcP}{{\mathcal P}}
\newcommand{\mcD}{{\mathcal D}}
\newcommand{\tsep}{t_{\mathrm{sep}}}
\newcommand{\tOp}{t_{\mathrm{op}}}
\newcommand{\vac}{{\mathrm{vac}}}

\def\SBUaffil{Stony Brook University, Stony Brook, NY 11794, USA}
\def\supportRBRC{\footnote{RHIC Physics Fellow Program of the RIKEN BNL Research Center}}
\def\RBRCaffil{RIKEN-BNL Research Center, Brookhaven National Lab, Upton, NY, 11973}
\def\Title#1{\begin{center} {\Large #1 } \end{center}}
\def\Author#1{\begin{center}{ \sc #1} \end{center}}
\def\Address#1{\begin{center}{ \it #1} \end{center}}

\newcommand\pubblock{\rightline{\begin{tabular}{l} \pubnumber\\
         \pubdate  \end{tabular}}}
\newenvironment{Abstract}{\begin{quotation}  }{\end{quotation}}
\newenvironment{Presented}{\begin{quotation} \begin{center} 
             PRESENTED AT\end{center}\bigskip 
      \begin{center}\begin{large}}{\end{large}\end{center} \end{quotation}}
\def\Acknowledgements{\bigskip  \bigskip \begin{center} \begin{large}
             \bf ACKNOWLEDGEMENTS \end{large}\end{center}}

\begin{document}
\begin{titlepage}
\pubblock

\vfill
\Title{Progress in the Nucleon Electric Dipole Moment Calculations in Lattice QCD}
\vfill
\Author{Sergey~Syritsyn\supportRBRC, \\in collaboration with Hiroshi~Ohki, Taku Izubuchi}
\Address{\RBRCaffil, \\ and \SBUaffil}
\vfill
\begin{Abstract}
Electric dipole moments (EDMs), which are sought as evidence of CP violation, require lattice
calculations to connect constraints from experiments to limits on the strong CP violation within
QCD or CP violation in new physics beyond the standard model.
Nucleon EDM calculations on a lattice are notoriously hard due to large statistical noise,
chiral symmetry violating effects, and potential mixing of the EDM and the anomalous
magnetic moment of the nucleon.
In this report, details of ongoing lattice calculations of proton and neutron EDMs induced by
the QCD $\theta$-term and the quark chromo-EDM, the lowest-dimension effective CP-violating
quark-gluon interaction are presented.
Our calculation employs chiral-symmetric fermion discretization.
An assessment of feasibility of nucleon EDM calculations at the physical point is discussed.
\end{Abstract}
\vfill
\begin{Presented}
Conference on Intersections of Particle and Nuclear Physics, \\
Palm Springs, CA, USA, May 29--June 3, 2018
\end{Presented}
\vfill
\end{titlepage}
\def\thefootnote{\fnsymbol{footnote}}
\setcounter{footnote}{0}
%

\section{Introduction}

Observing a non-zero nucleon electric dipole moment (nEDM) will be evidence of violation of
P,T-symmetries beyond the level of the Standard Model.
The latter alone is not sufficient to explain the observed excess of matter over antimatter in
the Universe.
Constraints from precise EDM measurements, which are projected to improve by two
orders of magnitude in the next decade, require knowledge of nucleon structure and
interactions to translate into bounds on BSM theories as well as the strong $\CP$-violation.
Such nucleon structure calculations are possible only with nonperturbative lattice QCD methods.

Interactions that can induce nucleon EDM have to be $P$- and $\CP$-odd, and can be ordered by
their dimension~\cite{Engel:2013lsa}
\begin{equation}
\mcL^{\CPbar} = \sum_i \frac{c_i}{\Lambda_{(i)}^{d_i-4}} \mcO_i^[d_i]\,,
\end{equation}
where $d_i$ and $\Lambda_{(i)}$ are the dimension and the scale of the (effective)
$\CP$-violating interactions. 
We use lattice QCD with chiral quark action to calculate the nEDM induced by 
the $d=4$ $\theta_\text{QCD}$-term,
as well as by the chromo-electric moment, the lowest-dimension ($d=5$) effective quark-gluon 
$\CP$-odd interaction that may be generated by extensions of the Standard Model:
\begin{align}
\nonumber
\mcL &= i \theta_\text{QCD} Q + i \sum_q \tilde\delta_q \mcC_q\,, 
\\
\label{eqn:Qtopo}
Q &= \frac1{16\pi^2} \sum_x\Tr \big[G_{\mu\nu}\tilde{G}_{\mu\nu}\big]_x\,,
\\
\label{eqn:cedm_def}
\mcC &= \bar{\psi}\,\big[\frac12 (G_{\mu\nu}\sigma_{\mu\nu})\,\gamma_5\big]\, \psi\,,
\end{align}
where we use the $su(N_c)$-algebra value for the gluon field strength related to the lattice
$a\times a$ plaquette as $U^P_{x,\mu\nu}\approx 1+ia^2G_{x,\mu\nu} +O(a^4)$ and to the notation
used in the perturbation theory $G_{\mu\nu} = g(G^{pert}_{\mu\nu})^a\lambda^a$, 
In addition, we calculate nEDM induced by the pseudoscalar density 
$\Tr[\lambda^a\lambda^b]=\frac12\delta^{ab}$.
\begin{equation}
\label{eqn:psc_def}
\mcP = \bar{\psi}\,\gamma_5 \, \psi\,,
\end{equation}
which are necessary for renormalizing the chromo-EDM operator $\mcC$.

\section{CP-odd nucleon structure on a lattice}

Calculation of nucleon electric dipole moments on a lattice can be done either from energy
shifts in a background electric
field~\cite{Aoki:1989rx,Shintani:2006xr,Shintani:2008nt,Abramczyk:2017oxr} or as the forward
limit of the $P,T$-odd electric dipole form factor
(EDFF)~\cite{Shintani:2005xg,Berruto:2005hg,Aoki:2008gv,Guo:2015tla,Shindler:2015aqa,
Alexandrou:2015spa,Shintani:2015vsx,Abramczyk:2017oxr},
\begin{equation}
\label{eqn:ff_cpviol}
\langle p^\prime,\sigma^\prime |J^\mu|p,\sigma\rangle_{\CPviol}
  = \bar{u}_{p^\prime,\sigma^\prime} \big[
    F_1(Q^2) \gamma^\mu 
    + \big(F_2(Q^2) + iF_3(Q^2)\big) \frac{i\sigma^{\mu\nu}q_\nu}{2M_N}
  \big] u_{p,\sigma}\,,
\end{equation}
where $Q^2=-q^2$ and $q=p^\prime-p$.
$\CP$-odd interactions that induce nEDM  can be introduced either as a finite-size modification
to the lattice QCD action~\cite{Aoki:2008gv,Guo:2015tla,Bhattacharya:2016oqm} 
\begin{equation}
S_\text{QCD}\to S_\text{QCD} + i \delta^\CPbar S = S_\text{QCD} + i\sum_{i,x} c_i [\mcO^{\CPbar}_i]_x\,,
\end{equation}
or as the first-order perturbation to nucleon correlation functions.
For example, the nucleon-current correlators in $\CPviol$ QCD vacuum are
\begin{gather}
\label{eqn:corr_cpviol}
\begin{aligned}
\langle N\,[\bar q \gamma^\mu q]\, \bar N \rangle_{\CPviol}
  &= \frac1Z\int\,\mcD U\,\mcD\bar\psi\mcD\psi e^{-S - i\delta^\CPbar S}
      \, N\,[\bar q \gamma^\mu q]\, \bar N \\
  &= C_{NJ\bar N} - i\sum_i c_i \,\delta^\CPbar_i C_{NJ\bar N}
    + O(c_\psi^2)\,,
\end{aligned}
\end{gather}
\\
where
$C_{NJ\bar N} = \langle N\,[\bar q \gamma^\mu q]\, \bar N\rangle$ and 
$\delta_i^\CPbar C_{NJ\bar N} = \langle N\,[\bar q \gamma^\mu q]\, \bar N \, 
  \sum_x [\mcO^\CPbar_i]_x \rangle$
are the nucleon-current correlation functions evaluated in the $\CP$-even QCD vacuum. 

In this work, we study the EDFF $F_3$ (see Eq.(\ref{eqn:ff_cpviol})).
Since the $F_3$ contribution vanishes from the matrix element~(\ref{eqn:ff_cpviol}) at $Q^2=0$, 
the nEDM cannot be computed directly from it and requires extrapolation $Q^2\to0$.
On the other hand, EDFFs also yields the Schiff moments $F_3^\prime(0)$ from their
$Q^2$-dependence.
To compute the matrix elements (\ref{eqn:ff_cpviol}) and extract the EDFF $F_3(Q^2)$, we
calculate the nucleon correlators
\begin{align}
\label{eqn:twopt_cpviol}
\{\delta^{\CPbar}\}C_{N\bar N}(\vec p,t) 
  &= \sum_{\vec x} e^{-i\vec p\cdot \vec x}\langle N_{\vec x,t} \bar{N}_{\vec0, 0}
    \, \{\delta^{\CPbar}S\}\rangle_\CPviol\,,\\
\label{eqn:threept_cpviol}
\{\delta^{\CPbar}\} C_{NJ\bar N}(\vec p^\prime,\tsep;\vec q,\tOp)
  &= \sum_{\vec y,\vec z} e^{-i\vec p^\prime\cdot \vec y+i\vec q\cdot\vec z}
              \langle N_{\vec y,\tsep} J^\mu_{\vec z,\tOp} \bar{N}_{\vec0,0}
  \, \{\delta^{\CPbar}S\}\rangle\,.
\end{align}
with and without insertions of the $\CP$-odd interactions.
More details on the analysis of the form factors can be found in a recent
paper~\cite{Abramczyk:2017oxr}.

$\CP$ violation modifies the nucleon-like states, as well as their overlaps with the
positive-parity nucleon interpolating fields on a lattice,
\begin{equation}
\label{eqn:alfive_def}
\langle vac |N|p\rangle \sim e^{i\alpha_5\gamma_5} u_p\,.
\end{equation}
where $\alpha_5$ is a ``parity-mixing angle'' and $u_p$ is the usual on-shell spinor.
Although this has been known since the original lattice calculation of EDFF~\cite{Shintani:2005xg}, 
the effect of this parity mixing on nucleon matrix elements in the $\CPviol$ vacuum has not been
correctly taken into account until Ref.~\cite{Abramczyk:2017oxr}. 
As a result, all previous references reported nucleon EDM values with a spurious contribution
from the nucleon anomalous magnetic moment,
\begin{equation}
\Delta d_n = -2 \alpha_5 \kappa_n
\end{equation}
and a similar contribution to the EDFF $\Delta F_3(Q^2)=-2\alpha_5 F_2(Q^2)$ from the Pauli form
factor $F_2$.
The effect of this correction was dramatic: all previously reported values of
$\theta_\text{QCD}$-induced 
nEDM~\cite{Shintani:2005xg,Berruto:2005hg,Aoki:2008gv,Guo:2015tla,Shindler:2015aqa,
Alexandrou:2015spa,Shintani:2015vsx}, once corrected, became consistent with zero within
uncertainties.
The mixing angle $\alpha_5$ is critical for correct determination of EDM on a lattice.
For example, the matrix element of the time component of the vector
current between nucleon states polarized in $\hat i$ direction is
\begin{equation}
\label{eqn:edm_from_J4}
\langle \vec p^\prime=0 | V^4|\vec p=-\vec q\rangle_{\CPviol} \propto \frac{q_i}{m} \big[
  (1+\tau) F_3(Q^2) + \alpha_5 G_E(Q^2) \big]
\end{equation}
where $\tau = Q^2 / (4 M_N^2)$, and $G_E$ is the Sachs electric form factor.
For the proton with $G_E(0)=1$, a biased value of $\alpha_5$ will lead to incorrect
determination of the proton EDM.

\begin{table}[htb]
\centering
\caption{
  Gauge ensembles used in this study. 
  The second column shows the action used and the reference where the ensemble was analyzed.
  \label{tab:gauge_ens}}
\begin{tabular}{ll|ccc|rr}
\hline\hline
$L_x^3\times L_t\times L_5$ & $S_F$[Ref] &
  $a\text{ [fm]}$ &   
  $m_\pi\,[\mathrm{MeV}]$ & $m_N\,[\mathrm{GeV}]$ &  
  Conf & Obsv.\\
\hline
$24^3\times64\times16$ & DWF\cite{Aoki:2010dy} &
  0.1105(6) & 340(2) & 1.178(10) & 
  1400 & 
  $\theta$-nEDM \\
$48^3\times96\times24$ & MDWF\cite{Blum:2014tka} & 
  0.1141(3) & 139.2(4) & 0.945(6) & 
  130 &  $\mcP$,$\mcC$-nEDM \\
\hline\hline
\end{tabular}
\end{table}

In this study, we use ensembles of QCD gauge configurations generated by the RBC/UKQCD
collaboration employing Iwasaki gauge action and $N_f=2+1$ dynamical chiral-symmetric
fermions with (M\"obius) domain wall action (see Tab.~\ref{tab:gauge_ens}).
One ensemble has unphysical heavy pion mass $m_\pi\approx340\text{ MeV}$ and is used to
study the $\theta_\text{QCD}$-induced nEDM.
The other ensemble has (very nearly) physical pion mass $m_\pi\approx139\text{ MeV}$ 
and is used primarily to calculate nucleon form factors and nEDMs induced by quark-gluon 
chromo-EDMs, and assess feasibility of $\theta_\text{QCD}$-induced nEDMs calculation
at the physical point.

\section{Nucleon EDM induced by the $\theta_\text{QCD}$-term}

Studying QCD $\theta$ term-induced effects is complicated by statistical noise due to
the global nature of the topological charge~(\ref{eqn:Qtopo}).
Its fluctuation $\delta Q=\sqrt{\langle Q^2\rangle} \propto \sqrt{V_4}$ grows with the lattice
volume $V_4$, and so does the statistical noise in the $\CP$-odd correlation 
functions~(\ref{eqn:twopt_cpviol},\ref{eqn:threept_cpviol}),
\begin{equation}
\label{eqn:cpodd_correction}
\delta^\theta\langle\mcO(x)\ldots\rangle = \theta\,\langle Q \mcO(x)\ldots\rangle\,,
\end{equation}
Since the correlation of the topological charge density decays as $\propto e^{-r/m_{\eta^\prime}}$ 
with distance $r$, it has been suggested that contributions to the topological charge $Q$
from points beyond $r\gg m^{-1}_{\eta^\prime}$ may be neglected in lattice
calculations of nEDM~\cite{Shintani:2015vsx,Liu:2017man}.
Such finite-range restriction may introduce a systematic bias in the computed value of
nEDM~(\ref{eqn:edm_from_J4}), if the ``effective'' parity mixing angle $\alpha_5$ is different
in the two-point nucleon~(\ref{eqn:twopt_cpviol}) and the tree-point
nucleon-current~(\ref{eqn:threept_cpviol}) correlation functions.
This difference may arise unless the nucleon sources and sinks in these $\CP$-odd Green's 
functions are constructed identically.
To illustrate this point, consider the transfer matrix formalism one relies upon
in order to obtain hadronic matrix elements on a lattice,
\begin{equation}
\label{eqn:c3pt_tm}
C_{3pt}^\mcO(\tsep,\tOp) = \sum_{n^\prime,n} 
  \langle\vac|N|n^\prime\rangle e^{-E^\prime_{n^\prime} (\tsep - \tOp)} 
  \langle n^\prime | \mcO | n \rangle
  e^{-E_{n} \tOp} \langle n | \bar{N} | \vac\rangle\,.
\end{equation}
If the $\CPviol$ interaction $\propto i\Tr[G\tilde{G}]$ is turned on at some moment $t<0$,
the QCD vacuum requires some Euclidean time $\Delta t$ to ``settle'' into the new $\CP$-violating
state $|\vac\rangle_{\CPviol}$.
In general, the nucleon field $\bar{N}$ acting on such transient QCD vacuum 
$|\vac\rangle\to|\vac\rangle_{\CPviol}$ will have Euclidean time-dependent overlap 
$\langle\tilde{n}|\bar{N}|\vac\rangle$ with the new nucleon-like states
\begin{equation}
|N^{(\pm)}\rangle \to |\tilde{N}^{(\pm)}\rangle 
  = |N^{(\pm)}\rangle \pm i\alfive |N^{(\mp)}\rangle\,.
\end{equation}
in the $\CPviol$ QCD vacuum.
Since apparent $\CPviol$ effects may arise from creation (annihilation) of the nucleon states by
the nucleon interpolating fields $\bar{N}$($N$) acting on the $\CPviol$ vacuum, as well as
evolution of these states in the $\CPviol$ vacuum, the sources and sinks in the $\CP$-odd
correlation functions~(\ref{eqn:twopt_cpviol},\ref{eqn:threept_cpviol}) must be constructed
identically to avoid ``fake'' $\CP$-violation.

\begin{figure}[htb]
\begin{minipage}{.4\textwidth}
  \centering
  \includegraphics[width=.65\textwidth]{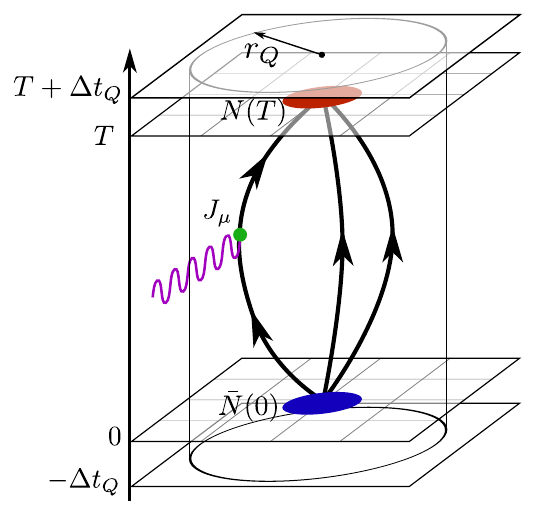}
\end{minipage}~\hspace{.05\textwidth}~
\begin{minipage}{.5\textwidth}
  \caption{
  Constrained sampling of the topological charge density~(\ref{eqn:Qtopo_cuts})
  for reducing the statistical noise in the
  $\CP$-odd three-point correlation function~(\ref{eqn:threept_cpviol}), 
  as well as the $\CP$-odd two-point correlation function~(\ref{eqn:twopt_cpviol}).
  \label{fig:qtopo_reduced}}
\end{minipage}
\end{figure}

To avoid this ambiguity, in our study we restrict the topological charge estimator separately in
time and space to a cylindrical volume $V_Q$ (Fig.~\ref{fig:qtopo_reduced}),
\begin{equation}
\label{eqn:Qtopo_cuts}
\tilde{Q}(\Delta t_Q,r_Q) 
  = \frac1{16\pi^2} \sum_{x\in V_Q}\Tr\big[ G_{\mu\nu} \tilde G_{\mu\nu}\big]_{x}\,,
\quad (\vec x,t)\in V_Q :\left\{\begin{array}{l} 
  |\vec x-\vec x_0|\le r_Q\,, \\
  t_0 - \Delta t_Q < t < t_0 + \tsep + \Delta t_Q\,,
\end{array}\right.
\end{equation}
where $t_0$ is the location of the nucleon source and $t_0+\tsep$ is the location of the nucleon
sink.
The $\CP$-odd correlation functions~(\ref{eqn:twopt_cpviol},\ref{eqn:threept_cpviol}) are
computed entirely inside the region~(\ref{eqn:Qtopo_cuts}) where $\CP$ violation is present
(i.e. where the reduced topological charge $\tilde{Q}$ is sampled).
The timelike cuts applied to $\tilde{Q}$ are symmetric with respect to the nucleon
sources and sinks and equal in the nucleon~(\ref{eqn:twopt_cpviol}) and
nucleon-current~(\ref{eqn:threept_cpviol}) correlation functions.
Additionally, we restrict $\tilde{Q}$ sampling in space to a 3D ball centered 
on the nucleon source, to further reduce the stochastic noise on large-volume lattices.
However, this restriction may interfere with the momentum projection 
in Eq.(\ref{eqn:threept_cpviol}) that requires summation over all $\vec y$ and $\vec z$.
\emph{We emphasize that convergence with $r_Q$ must be verified at
each momenta combination $p^\prime$ and $q$ to avoid bias}.

We use the lattice QCD ensemble with unphysical heavy pion mass $m_\pi\approx340\text{ MeV}$
(see Tab.~\ref{tab:gauge_ens}) to enhance the nEDM value in this preliminary study, since the
$\theta_\text{QCD}$-induced EDM decreases with $m_q\propto m_\pi^2$.
We calculate 64 low-precision and 1 high-precision samples using the \emph{AMA} 
sampling method~\cite{Shintani:2014vja}.
We analyze 1,400 gauge configurations separated by 5 MD steps to obtain 89,600 samples; samples
from each 10 MD steps (2 adjacent gauge configurations) are binned together.
The topological charge density in Eq.~(\ref{eqn:Qtopo_cuts}) is calculated from 
``5-loop-improved'' field strength tensor $G_{\mu\nu}$~\cite{deForcrand:1997esx}
computed from gradient-flowed~\cite{Luscher:2010iy,Luscher:2011bx,Luscher:2013vga} gauge
fields($\tau_{GF}=8a^2$).

\begin{figure}[htb]
\centering
\includegraphics[width=\textwidth]{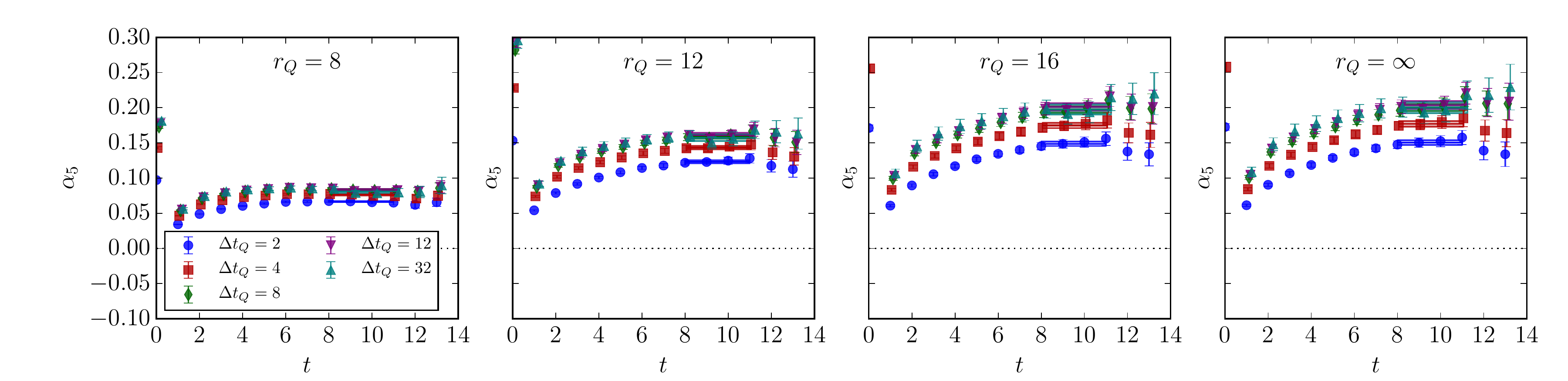}
\caption{
Effect of reduced $\tilde{Q}$ sampling on the nucleon parity mixing angle~(\ref{eqn:alfive_ratio}).
}
\label{fig:parmixing_dwf24c64}
\end{figure}

First we study the effect of reduced topological charge sampling on the mixing angle $\alpha_5$.
The mixing angle $\alpha_5$ is estimated with the $\{t,\Delta t_Q,r_Q\}$-dependent ratio
\begin{equation}
\label{eqn:alfive_ratio}
\alfivehat^{eff}(t) 
  = -\frac{\Tr\big[ T^+\gamma_5 \, \delta_{\tilde{Q}(\Delta t_Q,r_Q)}^\CPbar C_{N\bar N} (t)\big]}
          {\Tr\big[ T^+ \, C_{N\bar N}(t)\big]}
  \stackrel{t\to\infty} = \frac{\alfive}{\theta} \,.
\end{equation}
where $T^+=\frac{1+\gamma_4}2$ is the positive-parity projector.
Results for different values of $\Delta t_Q,\,r_Q$ are shown in
Fig.~\ref{fig:parmixing_dwf24c64}.
We generally observe convergence to the results obtained with the full topological charge 
Q~(\ref{eqn:Qtopo}) for $\Delta t_Q\gtrsim 8a$.
However, for the spatial cut $r_Q$ there is no convergence up to $r_Q\approx12a$, which is
$\approx52\%$ of the spatial volume.
We conclude that the lattice volume $V_3=(24a)^3\approx(2.7\text{ fm})^3$ 
is insufficient to benefit from the spatial cut $r_Q$, and should be explored with larger
spatial volumes.

\begin{figure}[htb]
\centering
\includegraphics[width=\textwidth]{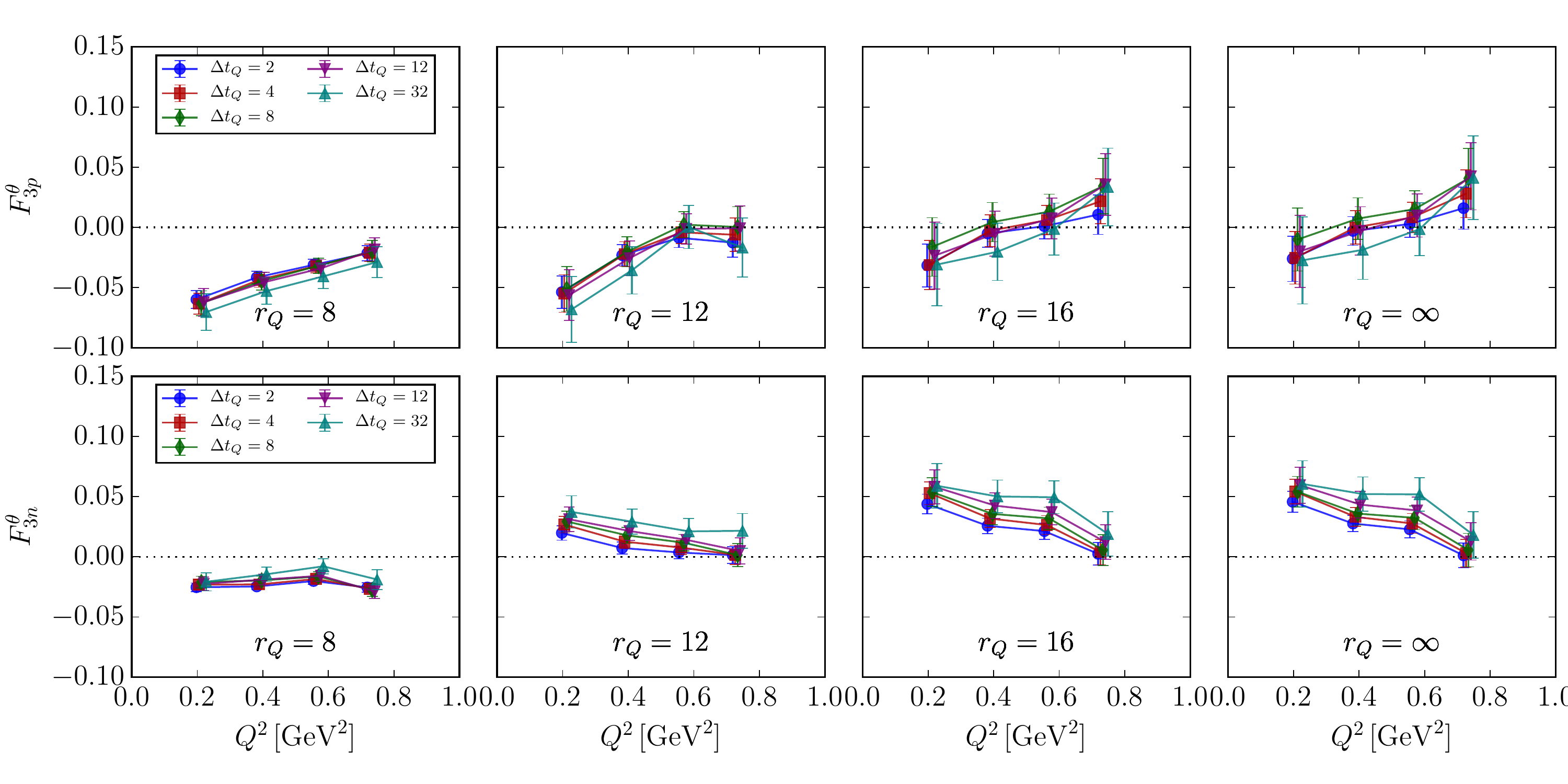}
\caption{
Proton and neutron electric dipole form factors induced by the $\theta_\text{QCD}$-term 
from lattice calculations with $m_\pi\approx340\text{ MeV}$ (only quark-connected contractions).
}
\label{fig:edff_dwf24c64_mpi330}
\end{figure}

The neutron and proton electric dipole form factors $\hat F^\theta_{3n,p}=F^\theta_{3n,p}/\theta$ 
computed for a range of $\Delta t_Q,r_Q$ are shown in Fig.\ref{fig:edff_dwf24c64_mpi330}. 
We compute only connected diagrams in this study.
The values for $\hat F^\theta_3$ are obtained using Eq.~(\ref{eqn:edm_from_J4}) with one value of
source-sink separation $\tsep=8a$.
Similarly to $\alfive$, we observe convergence for $\Delta t_Q\gtrsim8a$ but lack of convergence
for $r_Q\lesssim12a$.
Most importantly, we observe statistically significant value for the neutron $F_3$ even with the
full value of the topological charge $Q$, which has no bias from reduced sampling
$Q\to\tilde{Q}(\Delta t_Q, r_Q)$.
We can make \emph{a very preliminary ``ballpark'' estimate} for the value of $\hat
F^\theta_{3n}(0)\approx0.05$, without taking into account excited state effects or 
extrapolation $Q^2\to0$, \emph{which should be taken with a $100\%$ uncertainty.}
This value should only be used to check consistency with phenomenology and earlier lattice
QCD calculations.
For example, the \emph{corrected} value from calculations with Wilson
fermions~\cite{Guo:2015tla} constrains $|\hat F^\theta_3(0)|\lesssim0.06$ at a close value of
the pion mass $m_\pi\approx360\text{ MeV}$.
Leading-order extrapolation~\cite{Crewther:1979pi,Hockings:2005cn} 
$\hat d^\theta_n\propto m_{u,d}\propto m_\pi^2$ to the physical point yields values
\begin{equation}
\label{eqn:F3n_phys_estimate}
\hat F_{3n}^{\theta,\text{phys}}\approx0.01\,,
\quad\text{ or }
|\hat d^{\theta,\text{phys}}_{n}|
  = \frac{e}{2m_N} |\hat F_{3n}^{\theta,\text{phys}}|
  \approx0.001\,e\cdot\mathrm{fm}
\,,
\end{equation}
which is consistent with estimates from ChPT
and the QCD sum rules~\cite{Engel:2013lsa}.

\begin{figure}[htb]
\centering
\includegraphics[width=\textwidth]{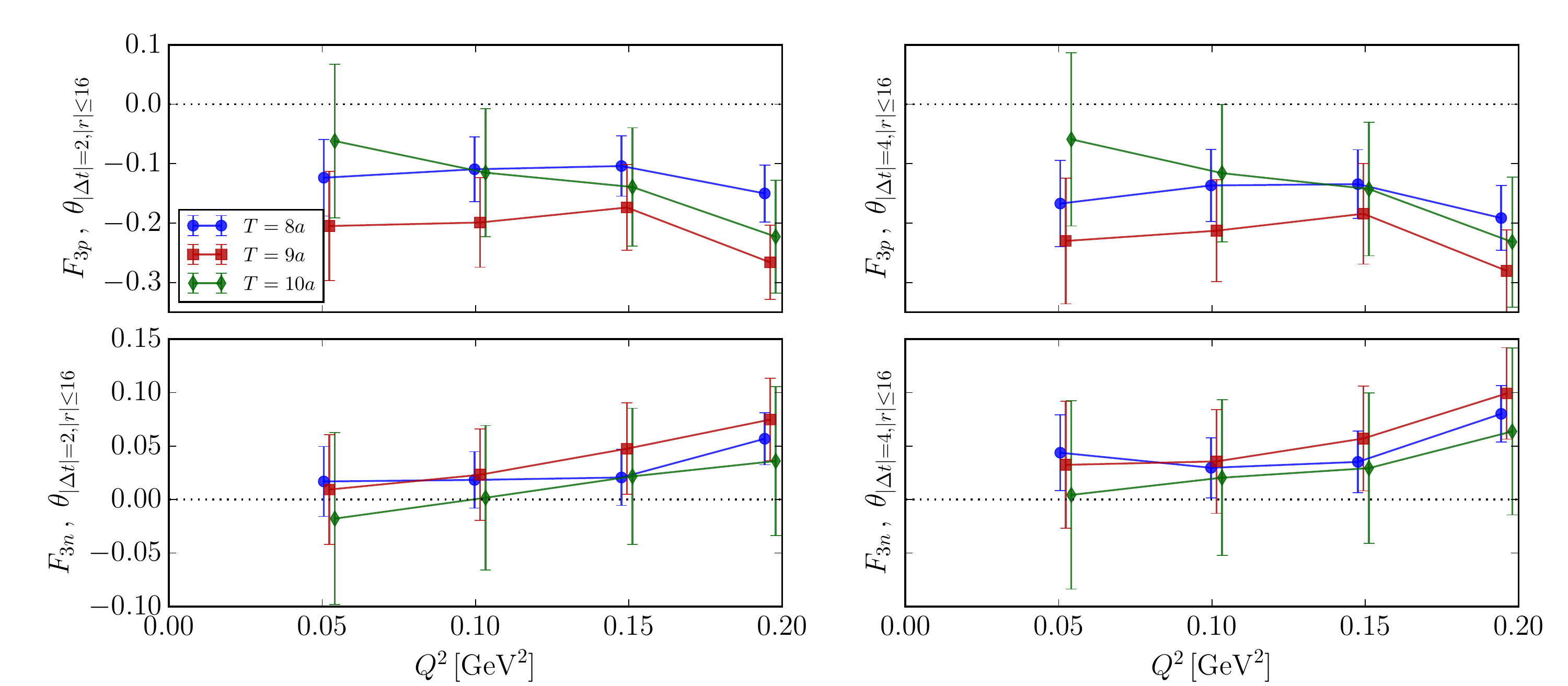}
\caption{
Proton and neutron electric dipole form factors induced by $\theta_\text{QCD}$-term 
from lattice calculations with physical quark masses.
}
\label{fig:edff_dwf48c96_mpiphys}
\end{figure}

Using our rough estimate for $\hat F^\theta_{3n}$, we can project the effort required for
computing nEDM at the physical point that is required to avoid model dependence due to
pion mass extrapolations $m_\pi \to m_\pi^\text{phys}$.
We have performed initial calculations using physical-quark ensembles with
$m_\pi\approx139\text{ MeV}$ (see Tab.~\ref{tab:gauge_ens}) with $\approx$ 33,000 statistical
samples and very aggressive time and space cuts in the topological charge estimator
$\tilde{Q}(\Delta t=2a,r_Q=16a)$.
We observe no signal for the neutron EDFFs (see Fig.~\ref{fig:edff_dwf48c96_mpiphys}), 
and the results are consistent with zero with the statistical uncertainty 
$\delta F_{3n}\approx0.05\ldots0.10$.
In comparison to the estimate~(\ref{eqn:F3n_phys_estimate}) above, we expect that the current
signal-to-noise ratio (SNR) $\approx0.01/0.05=0.2$ has to be improved by a factor of 5-10,
which requires $\times(25\ldots100)$ more statistics.
Alternative computing methods may have to be employed such as dynamical (imaginary)
$\theta^I$-term first explored in Ref.~\cite{Aoki:2008gv}.
Because nEDM calculations depend on contributions from non-trivial topological sectors,
dynamical $\theta^I$-term improves importance sampling for the EDM signal by inducing
$\langle Q\rangle\ne0$.
The dynamical $\theta^I$-term becomes more important at lighter pion masses, where 
light quarks suppress the fluctuation of topological charge.

\section{Nucleon EDM induced by quark chromo-EDM}

In this section, we report results from the ongoing calculations of nucleon EDM induced by the
dimension-5(6)\footnote{
  Quark-gluon chromo-EDM operator has dimension 6 above the electroweak scale 
  due to the Higgs field factor required by the electroweak symmetry.
}
chromo-electric quark-gluon interaction~(\ref{eqn:cedm_def}). 
The gluon field strength $G_{\mu\nu}$ in the chromo-EDM density~(\ref{eqn:cedm_def}) on a
lattice is computed with the ``clover'' form using $a\times a$ plaquettes,
\begin{equation}
\label{eqn:Gmunu_clover}
\begin{aligned}
\big[G_{\mu\nu}\big]_x^\text{clov} &= \frac1{8i} \big[
  ( U^P_{x,+\hat\mu,+\hat\nu} + U^P_{x,+\hat\nu,-\hat\mu} 
  + U^P_{x,-\hat\mu,-\hat\nu} + U^P_{x,-\hat\nu,+\hat\mu}) 
  - \mathrm{h.c.}\big]\,.
\end{aligned}
\end{equation}

\begin{figure}[htb]
\centering
\includegraphics[width=.8\textwidth]{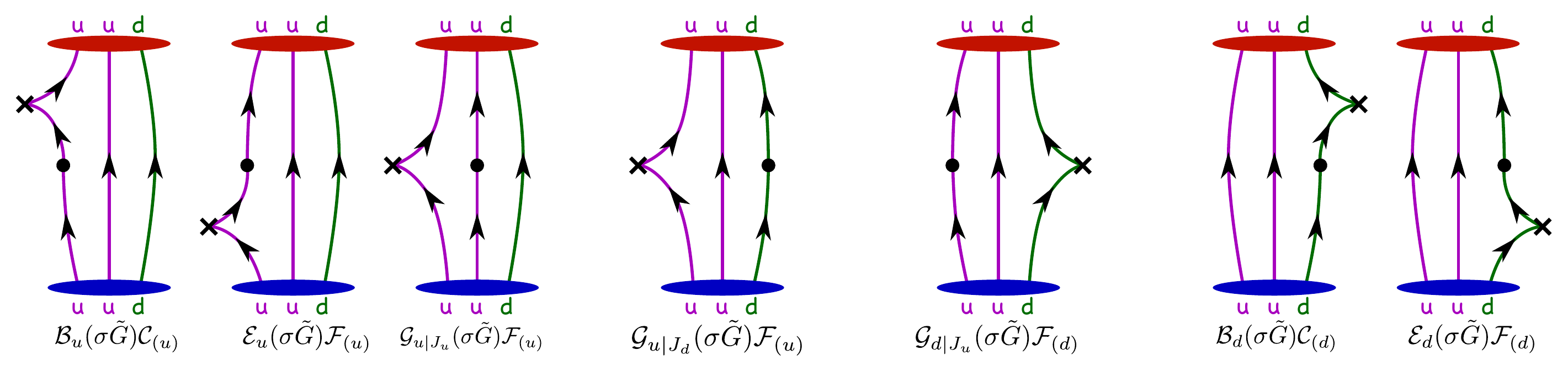}
\caption{
Connected lattice contractions for computing nucleon electric dipole form factors 
induced by chromo-EDM.
}
\label{fig:edff_diags}
\end{figure}

We evaluate only fully connected diagrams (see Fig.~\ref{fig:edff_diags}) for both the
$\CP$-even and -odd correlation functions~(\ref{eqn:twopt_cpviol},\ref{eqn:threept_cpviol}). 
More computationally demanding disconnected diagrams are required for a complete unbiased
calculation of isoscalar EDMs and effects of isoscalar quark chromo-EDMs, and will be
evaluated in the future.
We use the physical point ensemble (see Tab.~\ref{tab:gauge_ens}) and evaluate 256 low-precision
and 4 high-precision samples on each of 130 statistically-independent gauge configurations
separated by 40 MD steps, for the total of 33,280 statistical samples. 
All samples from the same gauge configuration are binned together.

\begin{figure}[htb]
\centering
\includegraphics[width=\textwidth]{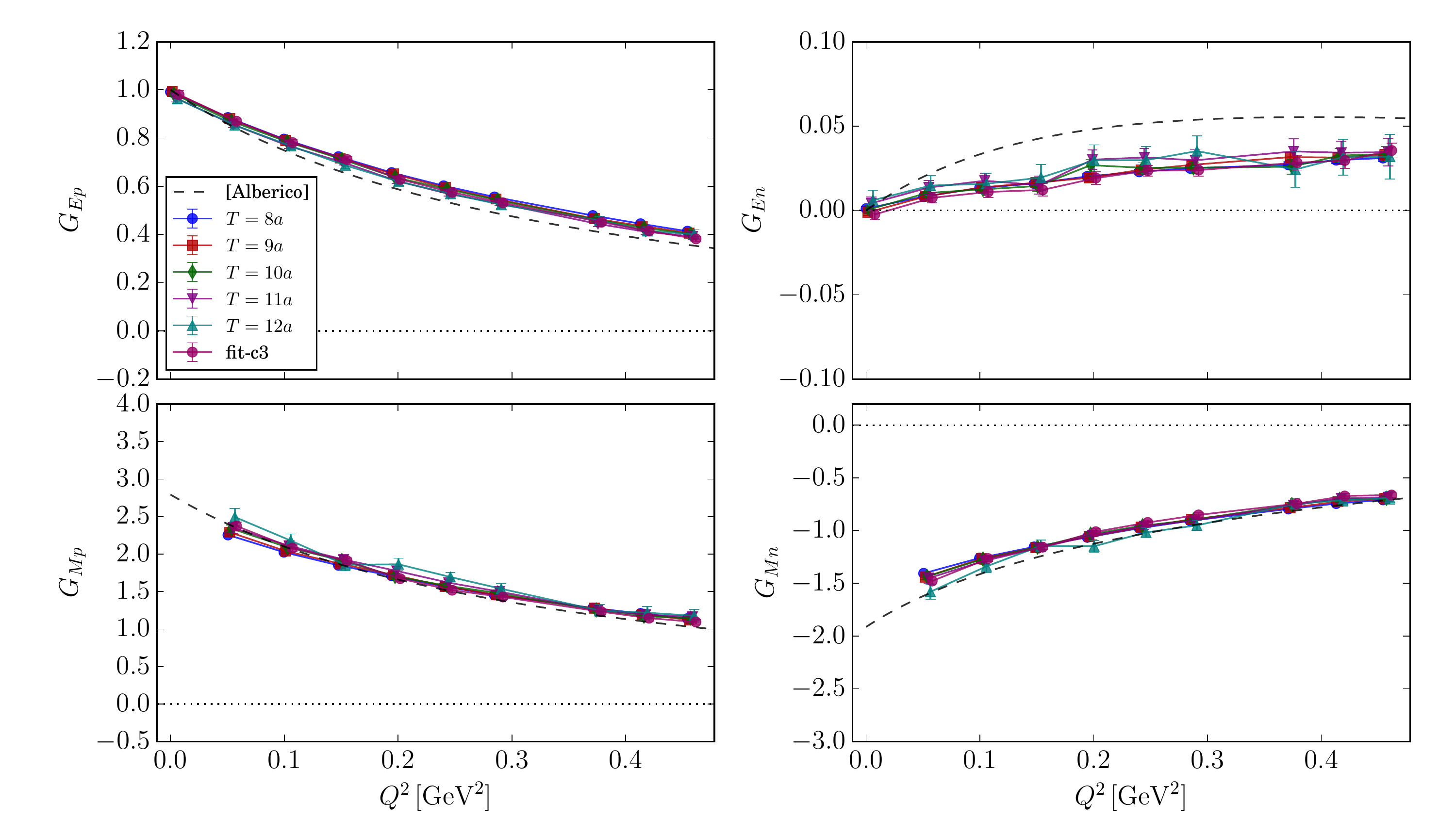}
\caption{
Nucleon electromagnetic form factors (connected contributions)
from lattice calculations with physical quark masses. 
}
\label{fig:ff_mpiphys}
\end{figure}

In Figure~\ref{fig:ff_mpiphys}, we show preliminary results for the proton and neutron
electromagnetic Sachs form factors $G_{E/M\,p/n}(Q^2)$ and their comparison the
phenomenological fits to experimental data~\cite{Alberico:2008sz}.
These form factors are extracted using the standard ``ratio'' method (see, e.g.,
Ref.~\cite{Hagler:2007xi}) for fixed source-sink separations $\tsep=(8\ldots12)a$ as well as
2-state fits using the state energies obtained from the nucleon two-point correlation functions.
Magnetic form factors $G_{Mp,n}$ show reasonable agreement with phenomenology. 
However, the electric form factors $G_{Ep,n}$ disagree for both the proton and the neutron,
which may be attributed to the missing contribution from the disconnected contractions.

\begin{figure}[htb]
\centering
\includegraphics[width=\textwidth]{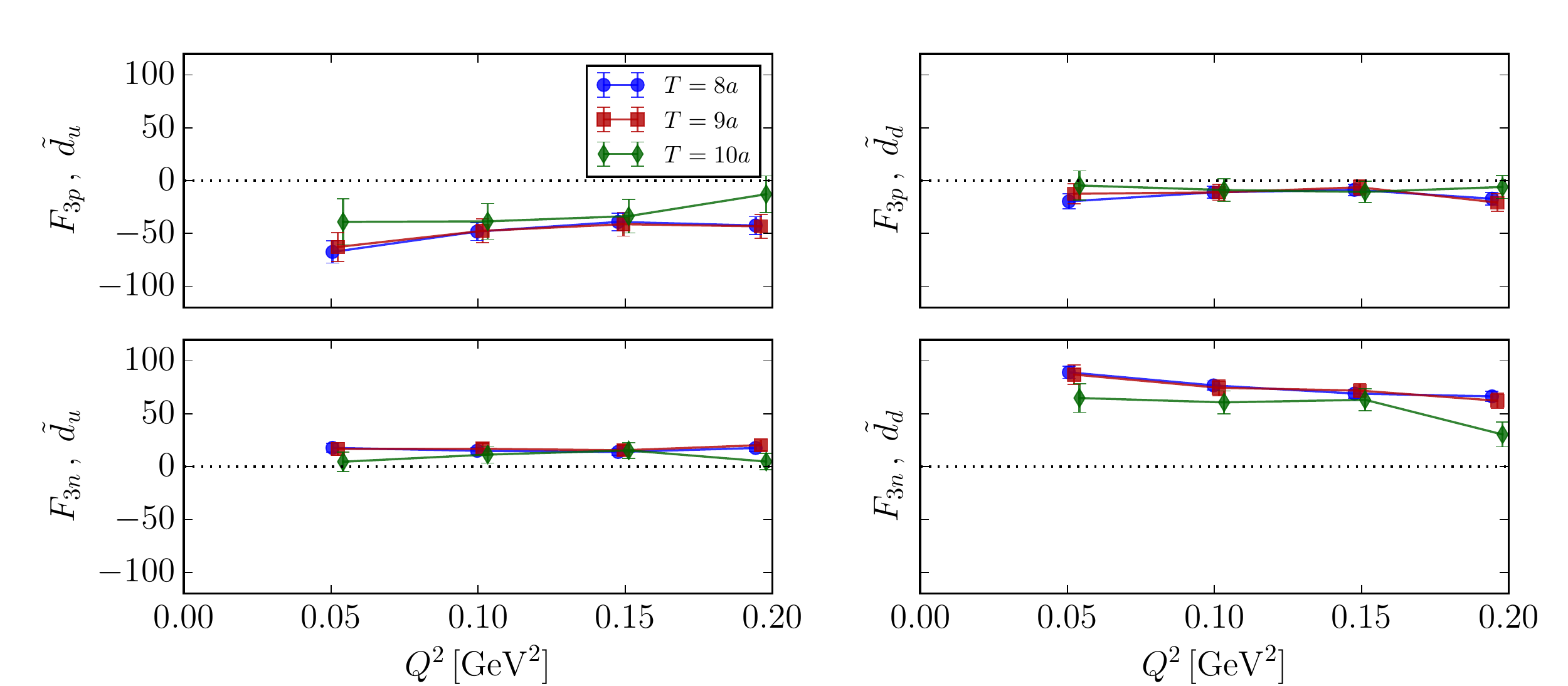}
\caption{
Proton and neutron electric dipole form factors induced by (lattice bare) chromo-EDM
from lattice calculations with physical quark masses. 
}
\label{fig:edff_cedm_mpiphys}
\end{figure}

\begin{figure}[htb]
\centering
\includegraphics[width=\textwidth]{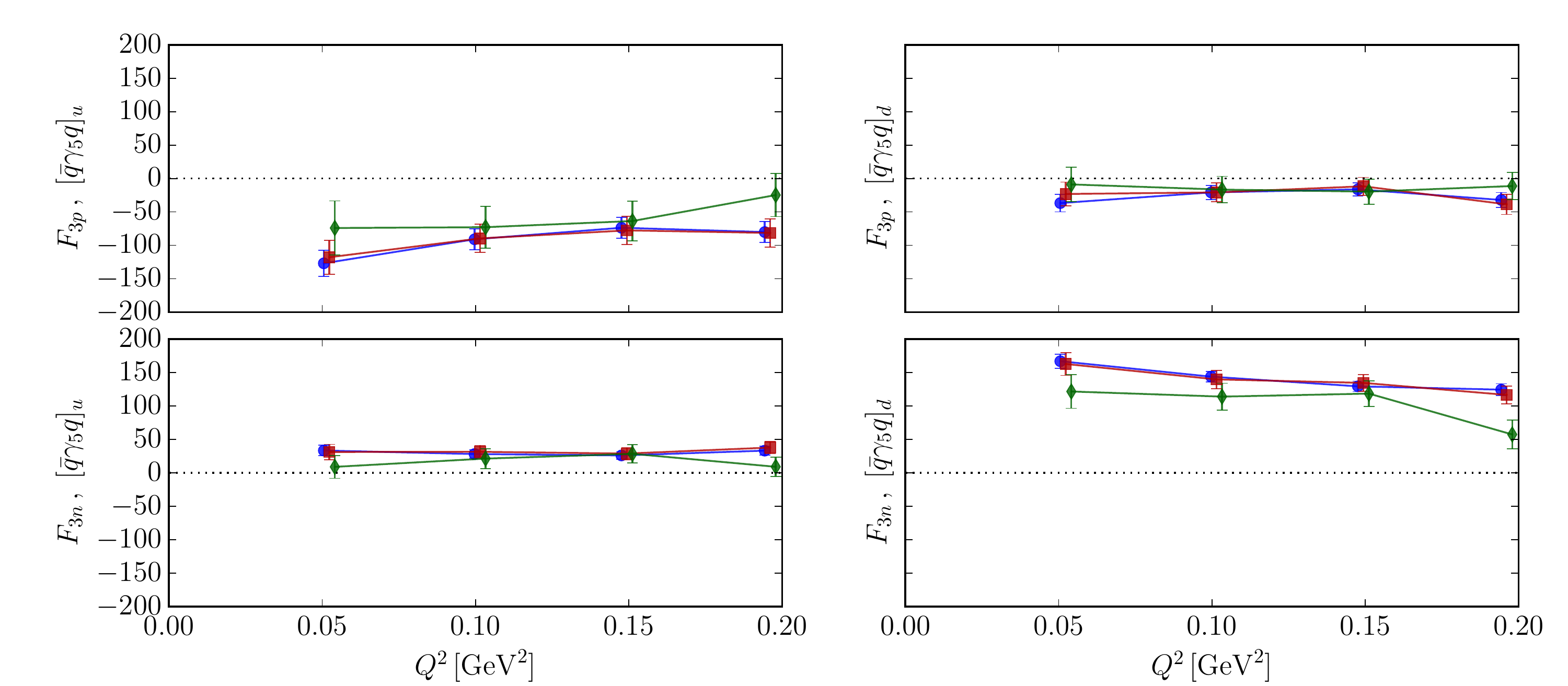}
\caption{
Nucleon electric dipole form factors induced by (lattice bare) pseudoscalar density operator 
required for renormalizing the chromo-EDM density operator at the physical point.
}
\label{fig:edff_psc_mpiphys}
\end{figure}

In Figures~\ref{fig:edff_cedm_mpiphys} and~\ref{fig:edff_psc_mpiphys} we show proton and neutron
EDFF induced by the \emph{unrenormalized (bare lattice)} quark chromo-EDM $\mcC$ and
pseudoscalar $\mcP$ densities.
These form factors are extracted using the ``ratio'' method with fixed source-sink separations
$\tsep=(8\ldots10)a$.
Data shows signal for both $\mcC$ and $\mcP$.
There is a peculiar dependence of nEDM on the flavor structure of $\CP$ violation:
the proton and the neutron EDMs are induced by the $\CP$ violation in the ``unpaired'' flavors,
i.e. in $u$- and $d$-quarks, respectively.

The final results for the chromo-EDM-induced nucleon EDMs requires renormalization that also 
has to be calculated nonperturbatively on a lattice. 
One proposed scheme is RI-SMOM, and perturbative matching to the $\MSbar$ scheme has been
calculated~\cite{Bhattacharya:2015rsa}.
Another approach is the position-space scheme~\cite{Gimenez:2004me,Chetyrkin:2010dx},
calculations of perturbative matching for which are underway.

\section{Summary and Outlook}
Calculations of nEDM on a lattice are important for interpreting constraints or results from 
nucleon and nuclei EDM measurements.
Ongoing calculations of nEDM induced by dim-5(6) quark-gluon $\CP$ violation show promising
results at the physical point.
However, their final precision will depend on renormalization that has not been computed yet,
and renormalized results may require substantially more statistics.
In contrast, calculations of $\theta_\text{QCD}$-induced nEDM at the physical point 
will be challenging and will require special techniques to tame the statistical noise caused by 
fluctuations of the global topological charge.
Direct calculations at the physical point may be at the limit of the current computing
capabilities, and one may have to use ChPT extrapolations of unphysical heavy-pion results.
Another approach is to simulate QCD with dynamical $\theta^I_\text{QCD}$ term to enhance
importance sampling for the $\CPviol$ observables.

\Acknowledgements
We are grateful for the gauge configurations provided by the RBC/UKQCD collaboration.
This research used resources of the Argonne Leadership Computing Facility, which is a DOE Office
of Science User Facility supported under Contract DE-AC02-06CH11357,
and Hokusai supercomputer of the RIKEN ACCC facility.
SS is supported by the RHIC Physics Fellow Program of the RIKEN BNL Research
Center.
TI is supported in part by US DOE Contract DESC0012704(BNL), and JSPS KAKENHI grant numbers
JP26400261, JP17H02906.
HO is supported in part by JSPS KAKENHI Grant Numbers 17K14309.

\end{document}